 \documentclass[final,3p,times]{elsarticle}

\usepackage{graphicx}
\usepackage{amssymb}

\usepackage[table,xcdraw]{xcolor}
\usepackage{fancyhdr}
\usepackage{microtype}
\usepackage{physics}
\usepackage{palatino}

\usepackage{titlesec}
\usepackage{sectsty}
\usepackage{graphicx}
\usepackage{wrapfig}
\usepackage{subfigmat}
\usepackage{fancyhdr}

\usepackage{caption}

\usepackage{lastpage}
\usepackage{wrapfig}
\usepackage{graphicx}
\usepackage{subfloat}

\usepackage{amsmath}
\usepackage{amssymb, amsfonts}

\usepackage{tikz}
\usepackage{verbatim}
\usetikzlibrary{arrows,shapes}
\def \bs {\,}

\tikzstyle{every picture}+=[remember picture]

\everymath{\displaystyle}

\journal{Journal of Nuclear Materials}

\begin{document}

\begin{frontmatter}


\title{Experimentally validated Multiphysics modeling of fracture induced by thermal shocks in Sintered UO$_2$ Pellets}




\author[1]{Levi D. McClenny\corref{Levi D. McClenny}\fnref{label_1}}
\author[2]{Moiz I. Butt\corref{Moiz I. Butt}\fnref{label_1}}
\author[2,4]{M. Gomaa Abdoelatef\corref{M. Gomaa Abdoelatef}\fnref{label_1}}
\author[2]{Michal J. Pate}
\author[2]{Kay L. Yee}
\author[3]{R. Harikrishnan}
\author[2]{Delia Perez-Nunez} 
\author[5]{W. Jiang}
\author[2]{Luis H. Ortega}
\author[2]{Sean M. McDeavitt}
\author[2,4]{Karim Ahmed\corref{Karim Ahmed}\fnref{label_2}}
\fntext[label_1]{Theses authors contributed equally to this study.}
\fntext[label_2]{corresponding author, Email address: karim.ahmed@tamu.edu}

\address[1]{Department of Electrical Engineering, Texas A$\&$M University, College Station, TX 77843, USA}
\address[2]{Department of Nuclear Engineering, Texas A$\&$M University, College Station, TX 77843, USA}

\address[3]{Department of Aerospace Engineering, Texas A$\&$M University, College Station, TX 77843, USA}
\address[4]{Department of Material Science and Engineering, Texas A$\&$M University, College Station, TX 77843, USA}

\address[5]{Computational Mechanics and Materials, Idaho National Laboratory, P.O. Box 1625, Idaho Falls, ID 83415-3840, USA}


\begin{abstract}
Commercial nuclear power plants extensively rely on fission energy from uranium dioxide (UO$_2$) fuel pellets that provide thermal energy; consequently, generating carbon-free power in current generation reactors. UO$_2$ fuel incurs damage and fractures during operation due to large thermal gradients that develop across the fuel pellet during normal operation. The underlying mechanisms by which these processes take place are still poorly understood. This work is a part of our combined experimental and computational effort for quantifying the UO$_2$ fuel fracture behavior induced by thermal shock. In this work, we describe an experimental study performed to understand the fuel fracturing behavior of sintered powder UO$_{2}$ pellets when exposed to thermal shock conditions, as well as a multiphysics phase-field fracture model which accurately predicts the experimental results. Parametric studies and sensitivity analysis are used to assess uncertainty. Experimental data was collected from multiple experiments by exposing UO$_{2}$ pellets to high-temperature conditions (900-1200$^{\circ}$C), which are subsequently quenched in sub-zero water. We exhibit that the fracture results gathered in the experimental setting can be consistently recreated by this work phase-field fracture model, demonstrating a reliable ability to our model in simulating the thermal shock gradients and subsequent fracture mechanics in the primary fuel source for Light-Water Reactors (LWRs). This model advanced the fundamental understanding of thermal shock and property correlations to utilize UO$_{2}$ fuel better.
\end{abstract}


\end{frontmatter}


\section{Introduction}

 Currently, all commercial nuclear power plants in US are light water reactors (LWRs), which use UO$_{2}$ pellets fuel rods. Despite downsides with swelling under irradiation and high-thermal strains due to poor thermal conductivity, UO$_{2}$ is still the primary choice for LWRs due to its high melting point (2865$^{\circ}$C), lack of phase change up to its melting point (unlike metallic fuels), good corrosion resistance, and fission product retention \cite{ortega2020thermal}. An important part of LWR operation is fracturing of the fuel as it fissions and heat that is eventually converted into electricity.
 
One major challenge associated with UO$_{2}$ during operation is its high temperature gradient yet low thermal conductivity. Since UO$_{2}$ has a very low thermal conductivity ~\cite{badry2019experimentally}, a large thermal gradient can build up (Soret effect) with a mass gradient leading to the initiation of pores. Formation of the high burn-up structure with fine grains and porous microstructure is a well-documented microstructural phenomena during nuclear fission~\cite{abdoelatef2019mesoscale, abdoelatef2019mesoscale2}. The high-density grain boundaries in a high burn-up structure (HBS) act as defect sinks and can reduce the concentration of point defects in its grain interior and improve its thermal conductivity as opposed to the as-fabricated dense, large-grain microstructure \cite{bai2016multiscale}. 

\begin{figure}[hbt]
    \centering
    \includegraphics[width=.5\linewidth]{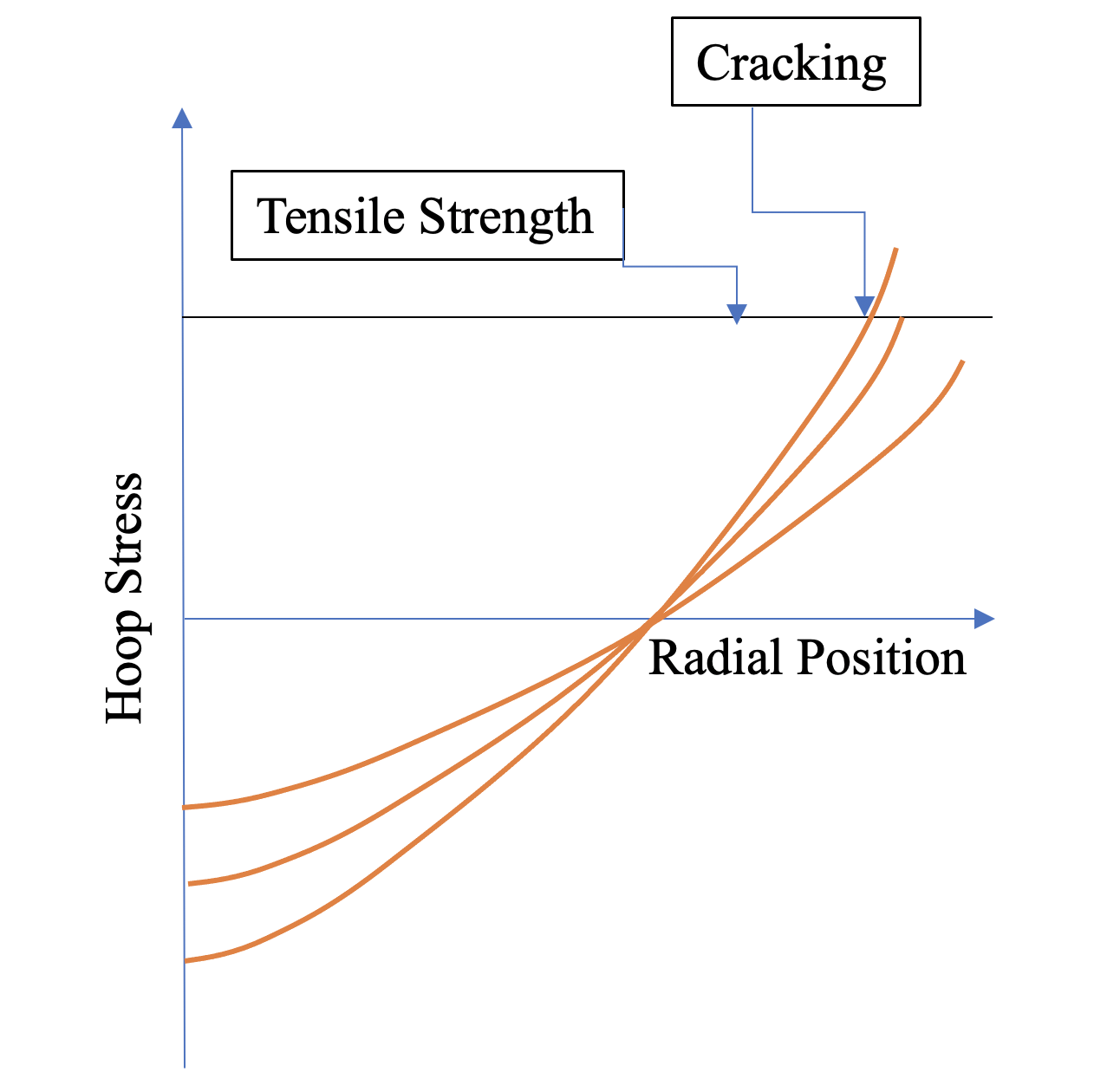}
    \caption{Plot of variation in hoop stress in fresh LWR fuel as a function of radial position during ramping to full power \cite{inlexperiments2019}.}
    \label{fig:hoop_stress}
\end{figure}

In as-fabricated unirradiated LWR fuel, volumetric heating from fission is mostly spatially uniform. As burnup increases, self-shielding effects result in higher volumetric heat generation around the perimeter of the fuel pellets. The heat is transferred radially outward from the fuel pellet, across the fuel-cladding gap, through the cladding, and finally into the coolant where it converts aqueous water into gaseous steam.
During reactor operation, nuclear fuels experience swelling and form voids and bubbles, which may cause the pellet to fracture at higher temperatures. Fuel fracturing may occur when the fuel is thermally shocked, or a rapid decrease in temperature as a result of the thermo-mechanical stress incurred by the grain boundaries of the material and varies depending on density, porosity, chemical composition, and microstructural state. Non-uniform thermal expansion in the fuel pellet causes compressive stresses in the center and tensile stresses on the exterior. This hoop stress as a function of radial position is pictured in Figure \ref{fig:hoop_stress}. It can be observed from the plot that radial cracking occurs as the tensile hoop stress surpasses the tensile strength of the fuel due to increasing power \cite{inlexperiments2019}. The tensile strength of UO$_{2}$ is about 150 MPa and fracture initiation is expected to occur when the hoop stress surpasses this value \cite{inlexperiments2019}. In this work, a method to simulate this fuel pellet behavior with a molten cartescal bath was carried out to develop a comprehensive model on the thermo-mechanical behavior of porous powder sintered UO$_{2}$ pellets subject to a substantial, rapid temperature decrease after being slowly heated to a temperature range (600-900 $^{\circ}$C), that is higher than the nominal operating LWR temperatures (300-400 $^{\circ}$C) to simulate accidents like transients. 

\section{Background}

Fuel fracturing in LWRs is a well expected occurrence during reactor operation, however, minimal experimental data with varying configurations (such as high-burnup fuel) of fracture initiation and propagation of UO$_{2}$ fuel pellets exists. Since replication of a LWR environment for a fuel pellet is difficult due to the temperature profile being a result of a combination of volumetric fission heating and coolant heat removal, few have taken on this challenge \cite{inlexperiments2019}. With this limited data availability, novel modeling approaches cannot be implemented with certainty. This work results would improve upon current literature data that allow further investigations for radial relocation, pellet-cladding mechanical interactions, and fragmentation phenomena to benefit fuel performance objectives in LWRs. To fulfill this requirement for further experimental validation, fuel fracturing experiments were carried out that subject the fuel to a temperature profile that allows for an observation of the cracking propagation and emphasizes fuel cracking rather than any unrelated phenomena. 

\begin{figure}[hbt]
    \centering
    \includegraphics[width=.5\linewidth]{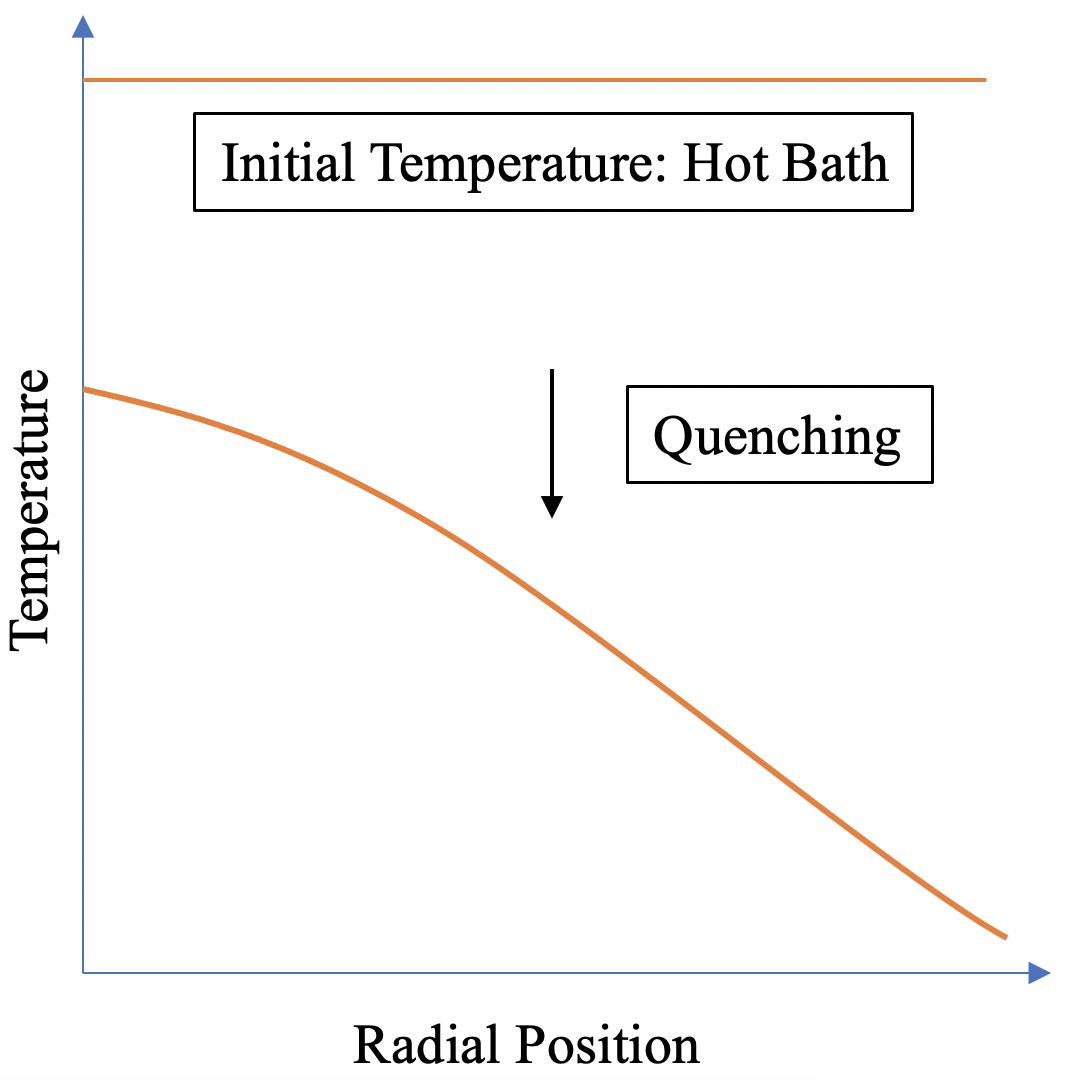}
    \caption{Plot of expected temperature profile during quenching after achieving a spatially uniform high temperature in the pellet \cite{inlexperiments2019}.}
    \label{fig:temp_radial_posi}
\end{figure}

As part of a three-component experimental study focused on different heating methods of UO$_{2}$ fuel, the purpose of this section was to slowly heat a fuel pellet contained in a metal tube assembly to a high temperature and subsequently quench said tube in a cold bath designed to insulate the top and bottom portions of the cylindrical pellet to emphasize radial heat flux \cite{inlexperiments2019}. Figure \ref{fig:temp_radial_posi} displays an expected temperature profile of the quenching process when cooling from a spatially uniform high temperature. The UO$_{2}$ pellets used for this study were manufactured at Texas A\&M University's Fuel Cycle and Materials Laboratory (FCML) using powder metallurgical methods and sintering to produce cylindrical UO$_{2}$ fuel pellets of approximately 9-10 mm in height and 10-11 mm diameter each. Fuel pellets of varying density, diameter, and heights were used in the thermal shocking experiments.

This contribution layout describes this work experiment method and procedure of the thermal-shock induced fracture in the UO$_{2}$ in section 3. This work model is explained in section 4, in which the model parameters are calculated based on the UO$_{2}$ thermal and mechanical properties available in the literature. Both experimental and modeling results were presented and discussed in sections 5 and 6, respectively. Finally, Section 7 summarizes the research conducted in this study and highlights the possible ways for further studies directions.

\section{Experimental Methods}\label{sec:methods} 

 Experimental data has been collected on fuel fracturing of sintered UO$_{2}$ pellets by thermally shocking them in high-temperature conditions (600-800$^{\circ}$C ternary carbonate salt) and subsequently in sub-zero conditions (-10$^{\circ}$C in water or 1:1 ethylene glycol:water). 
 More detail about the experimental procedures can be found in~\cite{ortega_paper}. Here, we present multiphysics computational modeling approaches and additional experimental results, extending the work performed in~\cite{ortega_paper} and adding supplemental discussion. 
The experimental set-up consists of a 
UO$_{2}$ pellet resting on a ceramic cylinder, which is inserted into a copper tube in a sealed Swagelok assembly. Insulation is placed on one side of the fuel pellet to increase thermal conductivity through the copper on one side. This tube assembly is constructed inside of a helium-filled glovebox in order to reduce the impact of oxidation upon the fuel pellet. A ‘successful’ experiment is one that manages to induce fracture in the pellet as a result of the sudden temperature decrease from the heating in the hot salt bath to quenching in the cold  bath. The thermal shock tests were characterized using two thermocouples attached to the assembly, one attached to the top of the pellet and one attached to the outside of the tubing to measure the temperature difference across the boundary. The inner thermocouple (0.3" diameter) is placed into a small divot drilled into the pellet and sealed with high temperature cement while the outer thermocouple is held in place using a hose clamp. Figure \ref{fig:data_temp_and_fracture} (b) shows an optical microscope image of a successfully cracked sample next to an example of temperature data from the same experiment, Figure \ref{fig:data_temp_and_fracture} (a). Note the rapid decrease in temperature as the sample is quenched in the cold water bath.
 
 In this study, as mentioned, several thermal shock experiments were conducted which were utilized for this work modelling validation, demonstrated in section~\ref{sec:res_comp}. 

\subsection{Experimental Procedure}
\begin{figure}[!hbt]
    \hfill
    \subfigure[Temperature data]{\includegraphics[width=9cm]{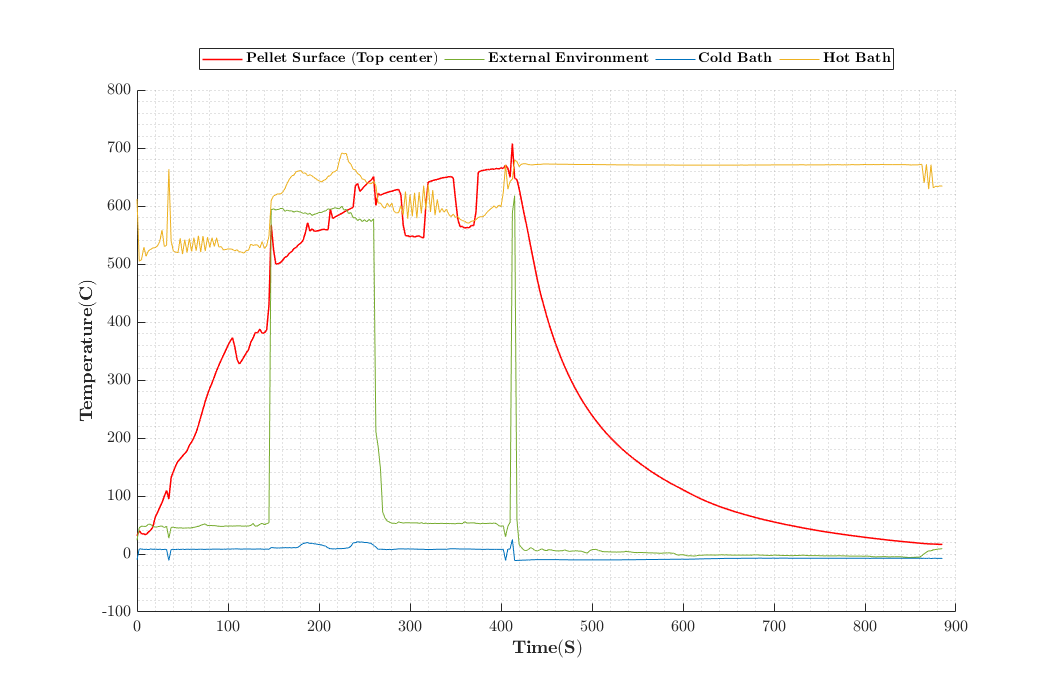}}
    \hfill
    \subfigure[Fractured pellet]{\includegraphics[width=6cm]{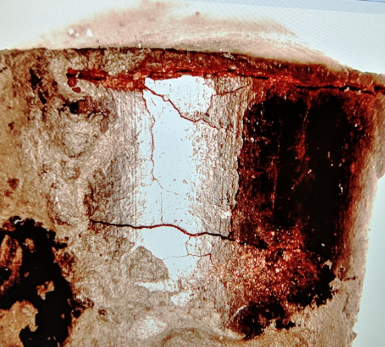}}
    \hfill
    \caption{(a) Temperature readings of thermocouples from a pellet submersion experiment of a UO$_{2}$ pellet (b) Hirox image of a successfully cracked UO$_{2}$ pellet}
    \label{fig:data_temp_and_fracture}
\end{figure}

The capsule was submerged in the molten salt bath until the pellet’s temperature reached equilibrium with that of the salt bath, about 11 minutes on average. A sample experimental temperature profile (which did result in fracture) collected via the LabVIEW software suite is shown in figure~\ref{fig:data_temp_and_fracture}(a). Upon reaching the desired temperature, the capsule was removed from the hot bath and immediately submerged into the cold bath. The capsule remained there until it cooled down to around room temperature. Table~\ref{tab:experiments} shows the times and temperatures for each experiment. For select experiments, temperature data for the heating process was inaccurate as depicted with an asterisk in Table ~\ref{tab:sinter}. However, upon removing the capsule, temperature readings stabilized with minimal variation. This malfunction in thermocouple readings may be caused by an interaction between the copper capsule and the molten salt bath. In the experiments that resulted in fracture which are primarily discussed in the main body of this contribution, physically compromised experiments were omitted from study.

The capsules were left in the fume hood for at least 24 hours before they were opened and the pellets were subsequently examined via aided and unaided means. Each pellet was imaged with an optical microscope for micro-level fracturing as well as unassisted user observation for macro-level fracture. The pellets that showed signs of cracking were placed into epoxy resin, sectioned, and polished for further imaging in the scanning electron microscope (SEM). SEM imaged experimental results are shown in sections~\ref{sec:images} and~\ref{sec:axial} for further study. More details about the experiment are included in the appendix.

\begin{table}[!ht]
\centering
\resizebox{\textwidth}{!}{\begin{tabular}{|c|c|c|c|c|c|c|c|}
\hline
 Pellet ID & \begin{tabular}[c]{@{}c@{}}Hot Bath \\ Temp (C)\end{tabular} & \begin{tabular}[c]{@{}c@{}}Density \\ (g/cc)\end{tabular} & \begin{tabular}[c]{@{}c@{}}Capsule Hot \\ Bath Time \\ (min)\end{tabular} & \begin{tabular}[c]{@{}c@{}}Highest Pellet \\ Temp (C)\end{tabular} & \begin{tabular}[c]{@{}c@{}}Cold Bath \\ Temp\end{tabular} & \begin{tabular}[c]{@{}c@{}}Capsule Cold \\ Bath Temp\end{tabular} & \begin{tabular}[c]{@{}c@{}}Lowest Pellet \\ Temperature\end{tabular} \\ \hline
 U6-71B    & 640               & 8.75           & 6                           & 640                     & *10            & 7                      & 35                        \\ \hline
 U6-40     & 670               & 9.82           & 7                           & 667                     & -11            & 8                      & 17                        \\ \hline
 U2*25     & 635               & 9.79           & 8                           & 600                     & -9             & 15                     & 0                         \\ \hline
 U4-53C    & 700               & 10.11          & 9                           & 662                     & -9             & 15                     & 0                         \\ \hline
 U4-42     & 565               & 10.31          & 10                          & 585                      & 4              & 10                     & 25                        \\ \hline
 U5-15A    & 683               & 10.38          & 11                          & 676                     & 5              & 15                     & 10                        \\ \hline
 U2-49     & 712               & 10.04          & 26                          & 669                     & 5              & 14                     & 10    
              \\ \hline
\end{tabular}}
\caption{The submersion times and temperatures of each experiment}
\label{tab:experiments}
\end{table}




\section{Multiphysics Modeling of Thermal Shock-Induced Fracture}\label{sec:model}
 This section briefly introduces a multiphysics modeling approach that couple heat transfer, mechanics, and fracture to simulate the fracture behaviour in the UO$_{2}$ pellet due to a thermal shock. Similar approaches have been conducted in literature, such as those in ~\cite{li2021multiphysics, bourdin2014morphogenesis}. However, the results of this work model are compared against the aforementioned experimental measurements to demonstrate the efficacy of our modeling approach in predicting the fracture propagation that is consistent with physical experimentation. The authors utilized the phase-field fracture model \cite{bourdin2008variational} implemented in the MOOSE framework \cite{gaston2014moose}, similar to Wen et al. \cite{jiang2020three} and Miehe et al. \cite{miehe2010thermodynamically, miehe2010phase}. Although this work is implementing phase-field fracture models, it is worth noting that the MOOSE framework is a multiphysics-capable simulation environment and is suitable for additional applications, including fluid dynamics simulations as recently evaluated by  \citet{ICONE28-64908}. Such multiphysics capabilities in MOOSE allow the pursuit of more detailed studies in the future.

In literature, there are several valuable techniques to numerically capture various fracture behaviors. Some of these successful methods are the discrete element method (DEM), the extended finite element method (X-FEM) \cite{babuvska1997partition, moes1999finite}, and the cohesive zone method (CZM) \cite{kamaya2007three, kamaya2009simulation}. Nevertheless, these methods are usually associated with high computational cost and the cracking interfaces are not directly simulated. In X-FEM, the cracks are typically defined as discrete discontinuities that ease the brittle fracture simulations in polycrystalline material \cite{sukumar2003brittle}, while the CZM approach applies a traction-separation method that indicates various fracture behaviors \cite{needleman1992micromechanical,ortiz1999finite}. Both methods (i.e., X-FEM and CZM) struggle to track the complex fracture evolution typically observed in real-world experimental data. Moreover, CZM shows mesh-dependent issues since the crack representations are usually limited to the boundaries of the elements \cite{jiang2020three}. 

In comparison, phase-field fracture modeling uses a length scale parameter that is assigned to control the damage band while the discrete fracture surfaces are represented through a diffused damage field. This method has improved simulation capabilities for complex crack patterns and topology, such as branching and coalescence \cite{karma2001phase, karma2004unsteady, henry2004dynamic, borden2012phase}. Furthermore, the phase-field model of fracture can explicitly track initiation of the crack interfaces through the evolution of a phase-field order parameter (i.e., the damage variable) by minimizing the total energy of a system composed of elastic and fracture energies. This approach was first applied in the late 1990s by Francfort \cite{francfort1998revisiting} and Bourdin \cite{bourdin2000numerical} and has been recently employed to investigate various fracture behaviors in brittle material \cite{zhang2020new, karma2004unsteady, borden2012phase} and nuclear materials \cite{chakraborty2016phase}. Additionally, it has also been implemented to investigate fracture behaviors in UO$_2$ \cite{clayton2015phase}, which is used in this study. 

UO$_2$ is a fluorite ceramic structure holding a cubic elastic symmetry. Strain energy is decomposed into two separate parts, tensile ( $\psi_{elastic}^+$) and compressive ( $\psi_{elastic}^-$), where only the former contributes to fracture. To implement this methodology, Strain Volumetric and Deviatoric Decomposition were used rather than other decomposition methods proposed, such as those in \cite{clayton2015phase,teichtmeister2017phase}, due to its superior performance for brittle materials. In addition, the strain spectral decomposition method, introduced by Miehe \textit{et al.} \cite{miehe2010phase, miehe2010thermodynamically} works well as demonstrated in \cite{jiang2020three}. We encourage readers to consult \cite{jiang2020three} for more details about the strain spectral decomposition scheme and \cite{gaston2014moose} for a more detailed explanation of the Strain Volumetric Decomposition utilized in this work. Consequently, in this model, the elastic energy will be released in a ``fracture energy'' manner, i.e.:

\begin{subequations}
\begin{align}
    F_{\text{total}} &= \Psi_{\text{elastic}} - \Psi_{\text{external}} + \Psi_{\text{fracture}}\\
    &= \Psi_{\text{elastic}} - \Psi_{\text{external}} + \mathcal{G}_c\int_{\Omega}\gamma_{\text{crack}} \text{ d}\Omega \label{eq_total_energy}
\end{align}
\end{subequations}

Where, $\Omega$ is the system domain, ($\mathcal{G}_c$) the critical energy release rate, $\gamma_{\text{crack}}$  is a geometrical discontinuous function to represent the crack topology, in which $\gamma_{\text{crack}}  = 1$ represents a fully cracked surface, and $\gamma_{\text{crack}}=0$ indicated intact (or ``unfractured'') material. In our phase-field fracture model, we utilize the order parameter $c \in [0,1]$ to parameterize this function, such that:

\begin{subequations}
\begin{align}
    \gamma_{\text{crack}} \approx \gamma(c;l) &= \frac{1}{2l} \left( c^2 + l^2 \left| \nabla c \right|^2 \right)\\
    \lim_{l\rightarrow 0} \gamma(c;l) &= \gamma_{\text{crack}} \label{eq: length scale going to zero}
\end{align}
\end{subequations}

Here $l$ is the model length-scale that ensures the diffusion of the crack width. To smooth this discontinuous function ($\gamma_{\text{crack}}$), the so-called ``degradation function'' is usually applied to the elastic energy in a way that Eq. 1 would be modified as:

\begin{subequations}
\begin{align}
    F_{\text{total}} &= \Psi_{\text{elastic}} - \Psi_{\text{external}} + \mathcal{G}_c\int_{\Omega}\frac{1}{2l} \left( c^2 + l^2 \left| \nabla c \right|^2 \right) \text{ d}\Omega\\
    &= \int_{\Omega} \psi_{\text{elastic}} \text{ d}\Omega - \int_{\partial\Omega} \psi_{\text{external}} \text{ d}\Gamma + \mathcal{G}_c\int_{\Omega}\frac{1}{2l} \left( c^2 + l^2 \left| \nabla c \right|^2 \right) \text{ d}\Omega\\
    &= \underbrace{\int_{\Omega} g(c)\psi_{\text{elastic}}^+ \text{ d}\Omega + \int_{\Omega} \psi_{\text{elastic}}^- \text{ d}\Omega}_{\text{degraded elastic energy}} - \underbrace{\int_{\partial\Omega} \bs{\tau} \cdot \bs{u} \text{ d}\Gamma}_{\text{external energy}} + \underbrace{\mathcal{G}_c\int_{\Omega}\frac{1}{2l} \left( c^2 + l^2 \left| \nabla c \right|^2 \right) \text{ d}\Omega}_{\text{fracture energy}} \label{eq_total_energy_modified}
\end{align}
\end{subequations}

Where, $\Gamma_{\text{crack}}$  is the crack set/surface. $g(c)$ is the degradation function that is works to degrade the elastic energy based on the local order parameter. In literature, there are many ways to form this degradation function. Here we utilize a type of degradation function similar to~\cite{jiang2020three}, thus:

\begin{align}
    g(c) = (1-c)^2(1-k)+k \label{eq_degradation}
\end{align}

To avoid numerical issues, we approximate an initial small value of  $k$ ($\approx 10^{-6}$) across the domain, and set $g(0)=1$ and $g(1)=0$. The total stress in the damaged material is defined as:

\begin{equation}
    \sigma = \frac{\partial F_{total}}{\partial \epsilon} = \left(\left(1-c\right)^2\left(1-k\right) + k\right) \frac{\partial \psi^+}{\partial \epsilon} + \frac{\partial \psi^-}{\partial \epsilon} = \left(\left(1-c\right)^2\left(1-k\right) + k\right) \sigma^+ + \sigma^-
\end{equation}

As mentioned above, in this work we are using Strain Volumetric and Deviatoric Decomposition method that describes the orthogonal decomposition of strain tensor in spherical and deviatoric components. Thus, the strain tensor $\epsilon$ is defined as follows: 

\begin{subequations}
\begin{align}
    \epsilon &= \epsilon_S + \epsilon_D \\
    \epsilon_D &= \frac{1}{n} \ tr\left(\epsilon\right)I 
\end{align}
\end{subequations}

where I denotes the n-dimensional identity tensor, also the strain energy defined as: 

\begin{subequations}
\begin{align}
    \psi^+ &= \frac{1}{2}\ \lambda \ \bigg \langle tr \left(\epsilon\right)^2 \bigg \rangle_+ + \ \mu \ \epsilon_D \cdot \epsilon_D \\
    \psi^- &= \frac{1}{2}\ \lambda \ \bigg \langle tr\left( \epsilon \right)^2 \bigg \rangle_-
\end{align}
\end{subequations}

Where $\lambda$ is the lame constant and $\mu$ is the shear modulus. The stress tensors are defined as 

\begin{subequations}
\begin{align}
    \boldsymbol{\sigma}^- & = \lambda \ \bigg \langle tr \left( \epsilon \right) \bigg \rangle \\
    \boldsymbol{\sigma}^+ &= \mathbb{C} \ \epsilon - \boldsymbol{\sigma}^-
\end{align}
\end{subequations}

where $\mathbb{C}$ is the fourth-order elasticity tensor. In this work we assume a quasi-static system, therefore the equilibrium equations can be obtained by taking the variations of the total internal energy $F_{total}$, then

\begin{equation}
\nabla \cdot \left[\left(\left(1-c\right)^2 + k \right) \sigma^+ + \sigma^- \right] = 0
\end{equation}

where $\sigma^+$ and $\sigma^-$ are the Cauchy stress described by~\cite{jiang2020three}. The damage variable field $c$ evolves to minimize the total free energy of the system according to the Allen-Cahn equation~\cite{allen1972ground}, thus

\begin{equation}
    \frac{\partial c}{\partial t} = -\boldsymbol{L}\frac{\partial F}{\partial c} = - \boldsymbol{L} \left( -2 \left (
 1-c\right) H_o^+ + \frac{\mathcal{G}_c}{l}c - \mathcal{G}_cl\nabla^2 c\right)
 \end{equation}

where $L$ is mobility and $H_0^+$ is a history varable to prevent crack healing. More detail on the strain-history functional can be found in Miehe \textit{et al.}~\cite{miehe_2010_p1, miehe_2010_p2} and Wen \textit{et. al}~\cite{jiang2020three}.  



In this work system analysis, the stress develops and evolves due to the thermal shock. Therefore, we couple the phase-field fracture model with the heat conduction equation so that the model can simulate fracture due to thermo-mechanical loading. The temperature variations are the key factors during the thermal sock analysis. The heat conduction is modeled as:

\begin{equation}
k_0\nabla \cdot \nabla T - \rho C_p\dot{T} + Q = 0
\end{equation}

where $k_0$ denotes the thermal conductivity, $T$ is the temperature $C_p$ is the heat capacity, $\rho$ is density, and $Q$ is the heat source. The internal stress due to the temperature variations is entirely dependent on the thermal expansion of the system. Thus, the stress tensor ($\sigma_0$) can be defined as 

\begin{equation}\label{sig_0}
    \sigma_0 = \mathbb{C} \left(\epsilon - \alpha\left( T - T_0\right)\right)
    \end{equation}

where $\mathbb{C}$ and $\epsilon$ are the aforementioned fourth-order elasticity tensor and the strain tensor, respectively. Additionally, $\alpha$ denotes the thermal expansion coefficient while $T_0$ is the reference temperature. Equations 9-12 are the governing equations for the system; solved simultaneously using the finite element method via the MOOSE framework~\cite{permann2020moose}.

\section{Experimental Results}\label{sec:res_exp}

\begin{figure}[!hbt]
    \centering
    \includegraphics[width=.2\linewidth]{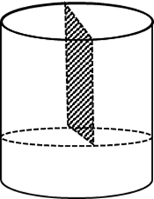}
    \caption{the pellets were cut into three parts which included two axial cuts and one radial cut to observe the fracture occurring in the samples}
    \label{fig:cross-section}
\end{figure}

\begin{figure}[hbt]
    \centering
    \includegraphics[width=.6\linewidth]{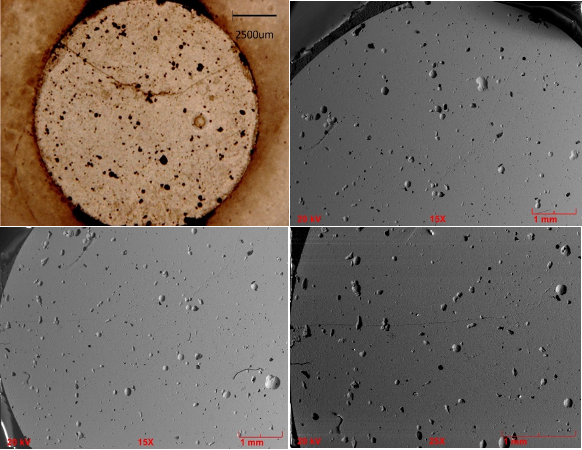}
    \caption{The experimental snapshots of various specimens microstructures reveal various porosity (number density and sizes) distributions.}
    \label{fig:number_density_and_sizes}
\end{figure}

\begin{figure}[!hbt]
    \centering
    \includegraphics[width=.9\linewidth]{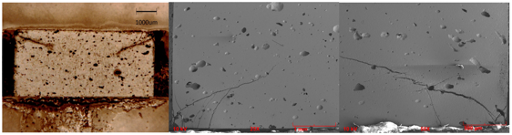}
    \caption{Experimental snapshots of various microstructures reveal various porosity (number density and sizes) distributions as well as subsequent fracture of the pellet. Some porosities are a few micrometers in size or even less, while bigger ones take ellipsoidal shapes. These larger pores have a size of several tens of micrometers.}
    \label{fig:pellet_res}
\end{figure}

The UO$_2$ samples were cut axially through the center and then radially through the middle, illustrated  in Figure \ref{fig:cross-section}. Numerous optical and SEM images are displayed in ~\ref{sec:appendix}, representing the final fracture results for each of the pellets, and can be characterized similarly to those shown in Figures \ref{fig:number_density_and_sizes} and ~\ref{fig:pellet_res}. The pellets were sintered in a mixture of argon and helium gas to prevent oxidation and bakeout was conducted for some pellets to remove any moisture as shown in ~\ref{sec:appendix}. The pellets were sintered to minimize porosity and the density of the pellets was determined using the Archimedes method also shown in Table ~\ref{tab:sinter}. The theoretical density of UO$_2$ is determined to be approximately 10.96 g/cc~\cite{fedotov2013theoretical}, leading to the density of the pellets being 80-90\% of the theoretical density. The variation in density is due to different sintering times, which leads to a change in porosity. A table of densities for various sintering experiments, the sintering processes used, as well as their subsequent quenching thermal quenching results is included in this work in ~\ref{sec:appendix}.

\begin{table}[!hbt]
\centering
\resizebox{.7\textwidth}{!}{\begin{tabular}{|c|c|c|c|}
\hline
Pellet ID & Mass (g) & Density (g/cc) & Sintering Profile                                                                                           \\ \hline
U6-71B    & 9.05     & 8.75           & 30 mins at 350C, 10 mins at 1800C                                                                           \\ \hline
U6-40     & 9.29     & 9.82           & \begin{tabular}[c]{@{}l@{}}30 mins at 350 C, 12 hours at 1700 C, \\  5 hours cool down to 25 C\end{tabular} \\ \hline
U2-25     & 9.41     & 9.79           & \begin{tabular}[c]{@{}l@{}}30 mins at 350 C, 8 hours at 1700 C, \\ 5 hours cool down to 25 C\end{tabular}   \\ \hline
U4-53C    & 9.08     & 10.11          & 24 hrs at 1790C                                                                                             \\ \hline
U4-42     & 9.16     & 10.31          & 325 bakeout, 18 hrs  at 1675C                                                                               \\ \hline
U5-15A    & 9.17     & 10.38          & 325 bakeout, 24 hrs at 1790C                                                                                \\ \hline
U2-49     & 9.11     & 10.04          & 325 bakeout, 3 hrs at 1700C \\ \hline                                   

\end{tabular}}

\caption{Sintering profiles and associated calculated densities for various experiments}
\label{tab:sinter}
\end{table}

\section{Model Results and Discussion}\label{sec:res_comp}

\begin{figure}[!ht]
\centering
    \hfill
    \subfigure[Experimental Observations]{\centering\includegraphics[width=.45\linewidth]{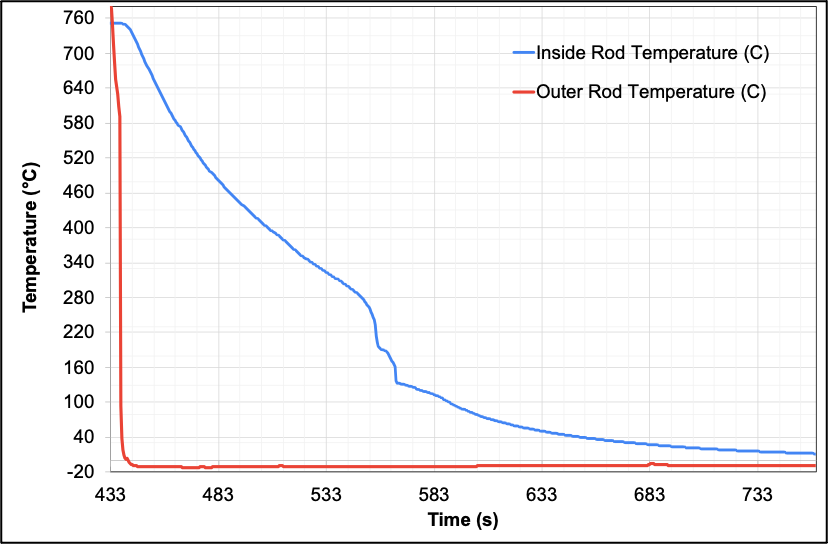}}
    \hfill
    \centering
    \subfigure[Model Simulations]{\centering\includegraphics[width=.48\linewidth]{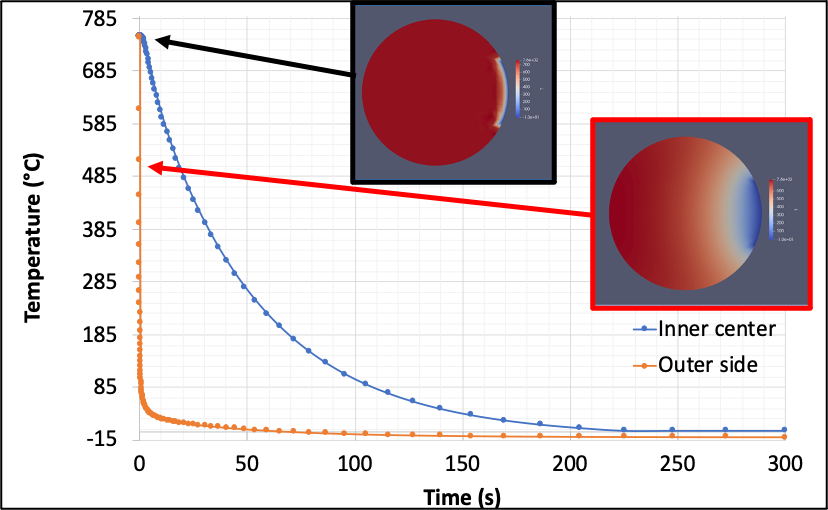}}
    \hfill
    \caption{The temperature evolution of the pellet outer surface and inner center points. Note that the counter timer in the experiment plot starts after the preheating process at approximately 430 seconds. The snapshots in the simulation plot show the temperature distribution at the initial and mid-time steps. The agreement of the temperature distribution and evolution between the model and the experiment results verified the basic model parameter selection.}
    \label{fig:sem}
\end{figure}

The phase-field fracture model described in section~\ref{sec:model} was implemented to investigate the fracture behavior of the UO$_2$ fuel pellet introduced to a thermal shock. The model was utilized to mimic the UO$_2$ thermal shock experiment as described in sections ~\ref{sec:methods} and ~\ref{sec:res_exp}. 
The fundamental model parameters were calculated based on the UO$_2$ thermomechanical properties, while the other effective parameters were determined based on a parametric study as described below. Due to the instantaneous temperature drop on the pellet surface, a large jump in the stresses throughout the pellet occurred. This behavior is mainly observed on the cooling side face. 
The rapid temperature change as well as the increased localized stresses are considered the main driving forces for the UO$_2$ fuel cracking. 
To obtain initial insights while minding computational costs, we only considered fresh fuel (zero burnup) in this model and then compared these simulated results with the previously described experiment outcomes. The model does not account for creep, i.e. fission gas release effects and grain growth influences were ignored, in the interest of creating a model that is computationally tractable with known and predictable physics. However, the qualitative effects of these phenomena should be considered in parametric study calculations of the energy release rate ($\mathcal{G}_c$) as illustrated throughout this section. The studied domain consists of a 3D fuel pellet with a 10 mm radius, and the axial direction height was set to 10 mm as well. The initial temperature condition of the pellet was set to 750 $^{\circ}$C. 
Only the determined ``contact region'' of the pellet was introduced to the low-temperature, -10$^{\circ}$C, simulating a thermal shock condition. In this study, the swelling and the densification consequence were considered negligibly small as described in~\cite{li2021multiphysics}. 
The model temperature evolution at the pellet center and right-side points against the experimental thermal shock data to verify the model capability to capture the fracture behavior was compared. This verification was presented in Figure~\ref{fig:sem}, which demonstrates consistency between the computational model and the experimental temperature behavior. 
Moreover, the model captured the temperature span correctly ($\approx$ 300$^{\circ}$C). Temperature span, in this context, implies the difference between the outer surface immediate drop and the gradual temperature drop at the centerline. 
Note that the counter timer in the experimental plot starts after the preheating process at nearly 430 seconds, as seen in Figure ~\ref{fig:data_temp_and_fracture} (a). The basic UO$_2$ thermomechanical properties utilized in this work are listed in Table ~\ref{tab:parameters}. 

\begin{table}[!ht]
\centering
\begin{tabular}{|l|c|c|c|c|}
\hline
Property         & Symbol & Value                        & Unit   & Reference \\ \hline
Youngs Modulus       &   $E$     & 358                          & GPa    &   ~\cite{jiang2020three, govers2007comparison}        \\
Poissons Ratio       &   $v$     & 0.23                         &        &     ~\cite{govers2007comparison}      \\
Thermal Conductivity &   $k$     & $\sim$5                      & W/m-K  &       ~\cite{badry2019experimentally}    \\
Fuel Density         &  $\rho$      & 10.97                        &  g/cm$^3$ & \\
Energy Release Rate  &  $g_c$       &   80                        &  MPa-mm &  This work \\
Length Scale         & $l$       & 1x10$^{-3}$ & mm     & This work \\
Viscosity            &   $\eta$     & 1x10$^{-8}$ & s/mm   &
~\cite{miehe_2010_p2} \\
\hline
\end{tabular}
\caption{Parameters utilized in this work for simulation results}
\label{tab:parameters}
\end{table}

\begin{table}[]
\centering
\begin{tabular}{c|c|c}
\hline
\multicolumn{1}{|c|}{\begin{tabular}[c]{@{}c@{}}Number of Test \\ Cases\end{tabular}} & \multicolumn{1}{c|}{\begin{tabular}[c]{@{}c@{}}Contact Area Fraction of the \\ Experiment Contact Area\end{tabular}}           & \multicolumn{1}{c|}{\begin{tabular}[c]{@{}c@{}}Energy Release Rate ($g_c$) range\\ (MPa-mm)\end{tabular}} \\ \hline
9   & Single Line      & 1.5, 5, 10, 30, 50, 80, 100, 150    \\
9     & 1/6              &          1.5, 5, 10, 30, 50, 80, 100, 150                                    \\
9      & 1/3     &  1.5, 5, 10, 30, 50, 80, 100, 150.   \\
9     & 2/3   &   1.5, 5, 10, 30, 50, 80, 100, 150 \\
9  & 1       &       1.5, 5, 10, 30, 50, 80, 100, 150       \\
\multicolumn{1}{c}{}   & \multicolumn{1}{l}{}    & \multicolumn{1}{l}{}     \\
 \rowcolor[HTML]{9DA6C2} \multicolumn{1}{c}{45}   & \multicolumn{1}{l}{Total Number of Test Cases}   & \multicolumn{1}{l}{}     \\
\rowcolor[HTML]{EDBF8F} \multicolumn{1}{l}{Selected Test Case}                        & \multicolumn{1}{l}{\begin{tabular}[c]{@{}l@{}}100\%, same as the \\ experiment test area\end{tabular}} & \multicolumn{1}{l}{80 Mpa-mm}

\end{tabular}
\caption{Detailed description of simulation test case script for the results provided in this section. 45 total simulations were run according to the above schedule.}
\label{tab:tests_exp}
\end{table}

It has previously been shown that the strength of brittle materials, such as UO$_2$, is strongly dependent on porosity microstructure~\cite{li2021multiphysics, jiang2020three}. Moreover, the porosity size and distribution may change with grain size variations. This trend is more pronounced for larger pore sizes~\cite{oguma1982microstructure}. The experimental work presented here obtained various microstructures containing micro and macro pores in different UO$_2$ specimens, as shown in Figures \ref{fig:number_density_and_sizes}, \ref{fig:pellet_res}, and the Appendix. 
In some of these specimens, many small spherical pores were almost uniformly distributed in the sample, whereas some exhibited ellipsoidal pores. 
Since the porosity of the UO$_2$ specimens undoubtedly affected the fracture mechanics in ways that are not completely understood, various simulations were conducted with different energy release rates ($\mathcal{G}_c$) until a value was settled upon that resulted in a fracture behavior similar to those found in the experiment. It is worth noting that similar methods were utilized experimentally by~\cite{doitrand2020uo2} and~\cite{henry2020fracture} to determine the strength and fracture toughness of UO$_2$. Here, a uniform distribution of porosity is assumed which may not actually be the case, but is still a reasonable assumption. A subsequent method could be developed in future works to account for the size, morphology, and distribution of porosity. 


A sensitivity study carried out in this work demonstrates an accordant prediction of the failure behavior and provides consistent information about material strength, fracture toughness, energy release rate, and other model parameters. The outcome of this sensitivity study, as presented in Table ~\ref{tab:tests_exp} and Figures \ref{fig:res1} and \ref{fig:res2}, is used to determine the corresponding model parameters by directly comparing various forms of the predicted failure behavior to the experimental measurements as well as to determine the energy release rate parameter. To that end,  various contact areas configurations and several contact areas of the cooling sides were considered. Some of these contact areas could have varied slightly from those seen in the physical experiments, but these various configurations improve our understanding of the ($\mathcal{G}_c$) parameter influence on the crack formations and evolutions, as well as account for some level of uncertainty in how large the contact region (cooling side area) truly was in the experiment. Additionally, variations on parameters such as ($\mathcal{G}_c$) and the contact region were intended to assist in capturing physical inconsistencies in the experiments, such as insulation, etc that were unaccounted for in the computational model. 
In all cases, we utilized a range of ($\mathcal{G}_c$) parameters along with the various contact area configurations. The details of the sensitivity study are summarized in Table~\ref{tab:tests_exp} and some of these simulations are shown with the remainder in~\ref{sec:appendix}.

\begin{figure}[!hbt]
    \centering
    \includegraphics[width=.7\linewidth]{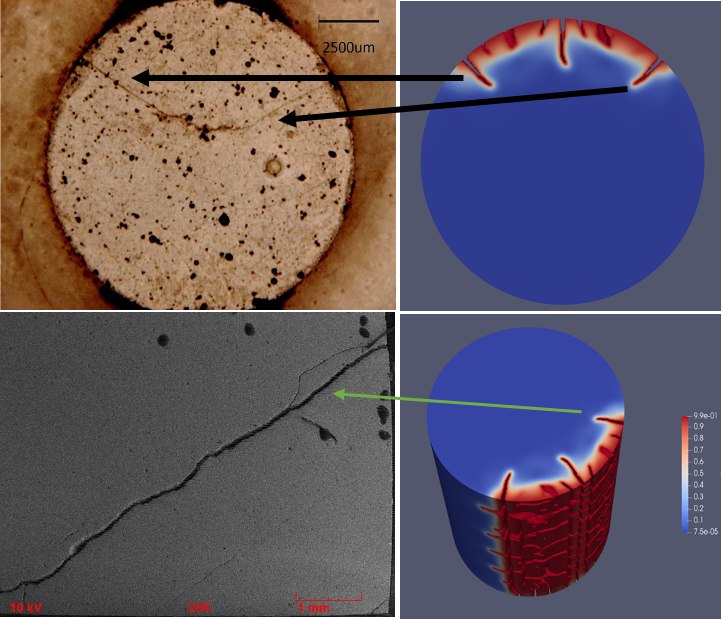}
    \caption{Comparison of final crack patterns induced by thermal shock between  this work experiment observations after the thermal shock testing (upper), and (lower) the elected model simulation (colored by the fractured order parameter field distribution). The model could capture the formation and evolution of primary radial cracks. As seen, there are two major (longer) radial cracks that formed immediately on the pellet circumferential boundary after the instantaneous drop in the outer temperature (thermal shock). Moreover, the simulated crack thickness was found to be similar to the experimental observation, as seen in left column. A minor differences between the model and the experiment results was noticed (such as the middle crack shown in the model simulations). These differences can be attributed to the absence of accounting for the effect of underlying microstructure (e.g., size and morphology of pores and grains) in the current model; overcoming such limitations is of future interest.}
    \label{fig:side_arrows}
\end{figure}

\begin{figure}[!hbt]
    \centering
    \includegraphics[width=.7\linewidth]{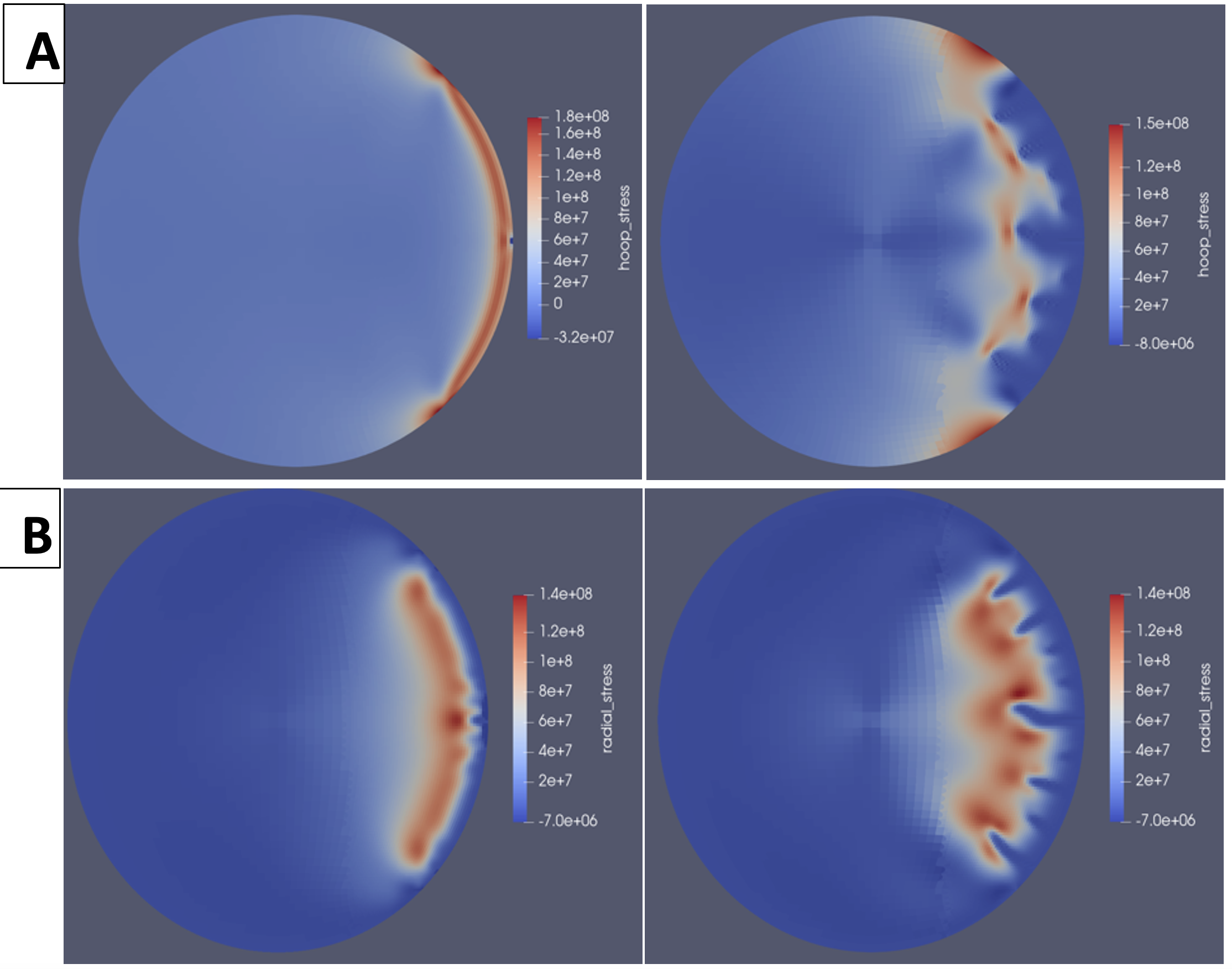}
    \caption{Snapshots illustrate the stresses formation and evolution before and after the cracks initiation and propagation for (upper) hoop stresses (lower) radial stress—the evaluation forms from left to right. The left column represents the stresses formation right after the instantaneous drop in temperature, while the right column shows the stresses evolution at the mid of the simulation time.}
    \label{fig:hoop}
\end{figure}

\begin{figure}[!hbt]
    \centering
    \includegraphics[width=.7\linewidth]{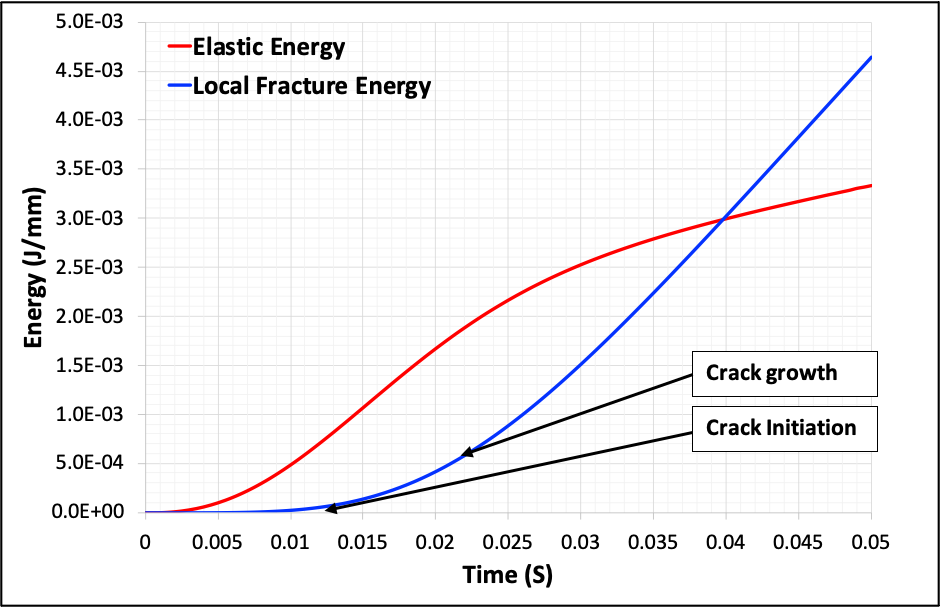}
    \caption{Evolution of elastic strain energy and the formation and growth of the local fracture energy based on fine mesh and length scale parameter of $10^{-3}$ mm for the elected simulation case; see the last row in Figure~\ref{fig:res2}. The elastic strain energy increases with increasing the temperature difference between the inner center and the outer surface (see Figure~\ref{fig:sem}). The cracks immediately initiated on the pellet outer surface at about $10^{-2}$ seconds, right after the instantaneous drop in temperature. The elastic strain energy drops quickly to generate new fracture faces.}
    \label{fig:energies}
\end{figure}

A demonstration of simulations based on what is perceived to be the approximate contact region as the experiment with different $\mathcal{G}_c$ values varying from 10 to 80 MPa$\cdot$mm is shown in figure~\ref{fig:res2}. 
Consistent results between the simulations and the experimental results with $\mathcal{G}_c$= 80 MPa$\cdot$mm was observed, as seen in the last row of Figure~\ref{fig:res2} and detailed in Figure~\ref{fig:side_arrows}. A reasonable agreement is found between the model prediction of the crack pattern and the thermal shock experiment results, as seen in Figure~\ref{fig:side_arrows}. 
The model could interpret the formation and evolution of such cracks induced by a thermal shock, particularly the primary radial cracks. 
There were two major (longer) radial cracks, as illustrated in the hoop and radial stress formation and evolution process in Figure~\ref{fig:hoop}. 
The cracks formed immediately on the pellet circumferential boundary after the instantaneous drop in the outer temperature (thermal shock). 
These results demonstrate consistent thermal elastic cracking behavior, wherein the radial cracks are generated by tensile hoop stress. Figure~\ref{fig:hoop} may be referenced for an improved illustration. 
It can be observed that the cracks do not propagate inward towards the center of the pellet. Such behavior is anticipated due to the existence of the compression zone that may hinder the cracks propagation, and a similar trend was observed by~\cite{li2021multiphysics}. 
Compared with the experiment where two major cracks merge in the middle, this phenomena could be attributed to the porosity of the microstructure causing inconsistent internal compression not anticipated by the computational model, resulting in the crack propagating entirely through the center of the pellet. 
It is worth noting that some of the tested specimens contain various porosity sizes and distributions and these uncertainties will have an impact on computational fracture predictions. For what is attributed to the same reason, it can be observed that a "middle" crack in the model resulted that did not propagate experimentally. 
Similarly, it was shown that the crack thickness is consistent with other samples with lower porosity. It is also worth noting that the model correctly reproduces different crack morphology's similar to the experimental outcomes. 
One motivation behind future works is to generalize the current model to account for microstructure porosity effects.
The elastic strain energy and fracture energy behaviors due to the thermal shock fracture were captured and presented in Figure~\ref{fig:energies}. As seen from this figure, the elastic strain energy rises with increasing the temperature difference between the inner and outer surfaces (see Figure~\ref{fig:sem}). As understood from this figure, the cracks immediately appear on the pellet outer surface at about $10^{-2}$ seconds right after the instantaneous drop in temperature. After some areas reached a full damage state (fully cracked areas), the elastic strain energy falls quickly as it is employed to generate new fracture faces, as described by Griffith~\cite{griffith1921vi}. Furthermore, the figure demonstrates that the local fracture energy or the dissipated energy initiated with the crack initiation and increases with further cracking growth. 
It is anticipated that this increase in the local fracture energy to be suppressed if the simulations ran long enough to capture the compressive loading in the central region as seen in ~\cite{li2021multiphysics}. 

\section{Conclusions and Future Work}

We introduced a new combined experimental and computational approach to investigate fracture in UO$_2$ pellets. First, the experimental method to promote thermal shock-induced fracture in UO$_2$ pellet is detailed. Second, the experimental data was utilized to parameterize and validate a multiphysics model capable of simulating and interpreting the experiments. Particularly, this work presents experimental approaches to inducing thermal shock-induced fracture in sintered UO$_2$ pellets and subsequent computational modeling. Currently, it is expensive to obtain experimental data to investigate fracture in sintered fuel pellets, and computational modeling of fracture of fuel pellets can lead to better predictions and understanding of this phenomenon. This work presents and discusses experimentally observed axial and radial fracture of UO$_2$ pellets and developed a multiphysics model based on the phase-field fracture techniques, which confirmed the thermal fracture physics at various microstructures conditions. Numerous experimental results were presented and utilized to validate our model which successfully captured the physics of thermal fracture for various mircrostructures. The model results are obtained by solving the coupled equations using the finite element method implemented in MOOSE. Validated via new experimental data, our model could explicitly, without a priori assumptions, simulate the initiation rate, sites, size, and morphology of the UO$_2$ pellets cracks induced by thermal shock. The model predictions demonstrate its capability to simulate the formation and the following growth of the cracks. The importance of the fuel pellets microstructure (e.g., porosity density) on the overall kinetics of the fracture formation and evolution were examined qualitatively. The model currently can only account for the effect of temperature on the kinetics of the evolution of the cracks. However, the model ignores a few factors of influence that are subject to future studies, such as the change of elastic and fracture properties with microstructure, temperature, and irradiation conditions. It is worth noting that this limitation can be alleviated by coupling the current phase-field model of fracture with the phase-field model of microstructure evolution of irradiation damage ~\cite{abdoelatef2019mesoscale, ahmed2020phase}. This also will be the subject of future work.

\section{Acknowledgements}
The experimental materials are based upon work supported by the U.S. Department of Energy NEUP IRP program under award agency number DE-NE-0008531. Any opinions, findings, and conclusions or recommendations expressed in this publication are those of the author(s) and do not necessarily reflect the views of the U.S. Department of Energy NEUP IRP program.
The modeling and simulation work is supported by the U.S. Department of Energy NEUP program under award agency number DE-NE0008979 at Texas A\&M University through a subcontract from the University of Wisconsin-Madison (0000001085)
Preliminary simulations of some of the results presented here were obtained as part of a term project for a computational materials science course (MSEN 619/ NUEN 660) offered at Texas A\&M University.  
Portions of this research made use of Idaho National Laboratory computing resources which are supported by the Office of Nuclear Energy of the U.S. Department of Energy and the Nuclear Science User Facilities under Contract No. DE-AC07-05ID14517.

\newpage
\appendix

\section{}
\label{sec:appendix}

\subsection{Experimental Setup}
A hot molten salt bath and a cold bath connected to a Julabo F-25 chiller were arranged in a fume hood for the experiments in the Fuel Cycle and Materials Lab at Texas A\&M University. The molten salt bath consisted of cartecsal (ternary carbonate) heat treating salt inside a crucible resting within a heater. The salt composition consists of lithium carbonate, potassium carbonate, and sodium carbonate. The hot bath had to be heated to a temperature above 432C to melt the cartescal. In order to reach the 600C degrees desired for the experiment, the heater for this bath was turned on three to four hours ahead of time. The cold and the hot bath were turned on at the same time but the cold bath reached the desired temperature of -10C to 5C much more rapidly. The cold bath contained a 1:1 mixture of ethylene glycol and water connected to the chiller with the same liquid as its working fluid. The liquid in the cold bath was replaced with water for the last three experiments since ethylene glycol can reduce convective heat transfer. Such behavior is undesirable for thermal shocking conditions. Both of the baths had type K thermocouples placed inside of them to record temperatures. 
\begin{figure}[!hbt]
    \centering
    \includegraphics{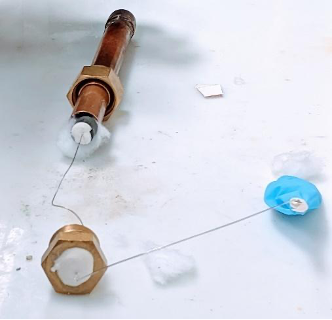}
    \includegraphics[width=.37\linewidth]{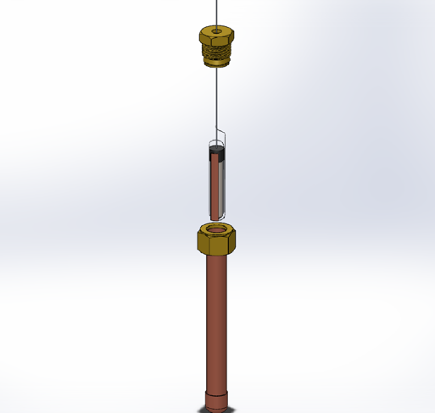}
    \caption{(left) Insertion of insulation and pellet U6-71B attached to the thermocouple into the capsule. (right) Solidworks rendering of the placement of the pellet in the capsule.}
    \label{fig:cap}
\end{figure}

The UO$_2$ pellets were placed inside a capsule, with a helium environment to prevent oxidation, for each experiment. The capsule consisted of copper tubing, an end cap, and brass fittings used to seal the top while allowing any sensors to go through, shown in Figure~\ref{fig:cap}. For the majority of the experiments, the capsule was a 177.8 mm long copper tube with a diameter of 15.88 mm (5/8in), a copper brazed end cap on the bottom, and a Swagelok brass fitting on the top. The Swagelok fitting had a small hole drilled through the top for the UO$_2$ thermocouple. Resbond 940HE epoxy held the wires in place and the top was sealed to prevent leakage of coolant mixture into the capsule and interacting with the pellets.

Inside the capsule, a piece of alumina of a diameter similar to the UO$_2$ pellets (typically 12.7 mm or 1/2 in) was first placed to eliminate axial thermal contact between the bottom of the capsule and the UO$_2$ so that conductive radial heat transfer was the primary method of heat transfer to occur. This was intentionally designed to form a stress concentration on the contact region to induce fracturing. For most of the experiments, the UO$_2$ pellet sat on top of the alumina and insulation was placed between the wall of the tube and the pellet. This insulation served to push the pellet to be in contact with one side of the capsule. Figure~\ref{fig:cap}\textit{(l)} shows an example of the insulation as it is being placed inside the capsule. 
It was placed in such a way that the foil formed thermal contact with the tubing walls on two opposite sides of the pellet. Figure~\ref{fig:cap}\textit{(r)} shows the SOLIDWORKS model of this setup. A type K thermocouple was placed into direct contact with the top center of the pellet and epoxied in place with the Resbond 940HE after being threaded through the hole in the Swagelok fitting. For the last four experiments, a small divot was drilled into the top of the pellet to improve the thermocouples placement on the pellets centerline. A LUNA strain sensor was also attached to the pellet sides using Durabond 952 and threaded through the top for the U6-40 experiment.

\begin{figure}[!hbt]
    \centering
    \includegraphics[width=.8\linewidth]{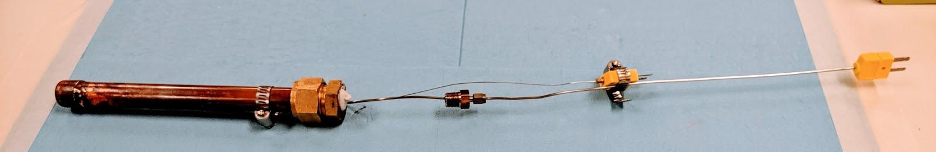}
    \caption{A finished capsule before it was submerged.}
    \label{fig:cap_fin}
\end{figure}

After the sensors were attached to the pellet and the Swagelok top fitting, the capsule was assembled and sealed inside a helium filled glovebox to prevent pellet oxidation. 
Another type K thermocouple was attached to the outside of the capsule using a hose clamp, as shown in Figure~\ref{fig:cap_fin}. The two thermocouples attached to the inside and outside of the capsule and the thermocouples in each of the baths were all connected to a NI-cDAQ 9171 measurement chassis recording all temperatures using a custom LabVIEW program on a laptop.

\newpage
\subsection{Additional Results}
\subsubsection{Simulation Results}

\begin{figure}[!hbt]
    \centering
    \includegraphics[width=.7\linewidth]{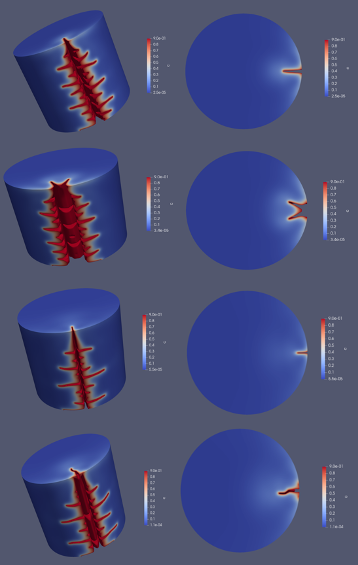}
    \caption{Snapshots of selected test cases simulations present the effect of different $g_c$ and different contact areas on the formation and evolutions of the cracks induced by thermal shock in UO2 fuel pellets. The first two rows show the cracks behaviors based on $g_c=10$ MPa-mm and the last two rows at $g_c =10$ MPa-mm. The first and third rows are simply a single line contact area, while the second and fourth ones were reproducing a wider contact area, almost 1/6 of the actual experiment contact area.}
    \label{fig:res1}
\end{figure}

\newpage
\begin{figure}[!hbt]
    \centering
    \includegraphics[width=.7\linewidth]{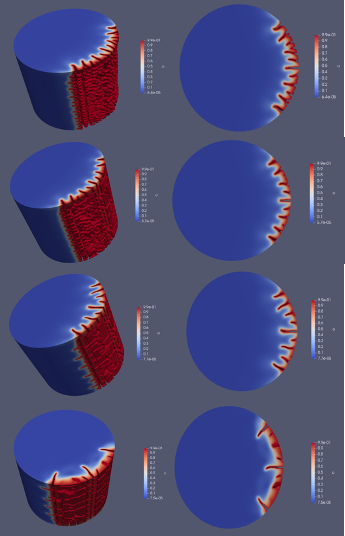}
    \caption{Snapshots of selected test cases simulations manifest the effect of  $g_c$ values, with the case of actual contact area, on the formation and evolutions of the cracks induced by thermal shock in UO2 fuel pellet. The damage shape evolutions with  $g_c=10$ MPa-mm was seen in the first row. While the elected evolution with  $g_c=80$ MPa-mm, which was found to be well-matched with the experimental results, is presented in the last row. The second and third rows show the simulations with $g_c=30$ MPa-mm and  $g_c=50$ MPa-mm, respectively.}
    \label{fig:res2}
\end{figure}

\newpage
\subsubsection{Radial Fracture Images}\label{sec:images}

\begin{figure}[!hbt]
    \centering
    \includegraphics[width=.6\linewidth]{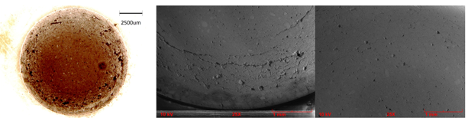}
    \caption{The experimental snapshots of U66-71B radial cut microstructure reveal various porosity (number density and sizes) distributions. Some porosities have a few micrometers size, or even less, while bigger ones take ellipsoidal shapes. These larger pores have a size of several tens of micrometers. A more detailed investigation of the effect of porosity microstructure on fracture formation and evolution is a part of future studies.}
    \label{fig:temp}
\end{figure}

\begin{figure}[!hbt]
    \centering
    \includegraphics[width=.6\linewidth]{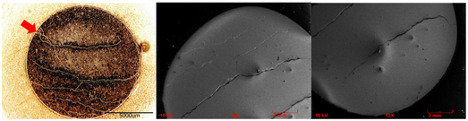}
    \caption{The experimental snapshots of U5-15A radial cut microstructures.}
    \label{fig:temp}
\end{figure}

\begin{figure}[!hbt]
    \centering
    \includegraphics[width=.6\linewidth]{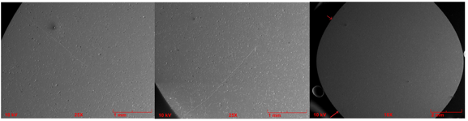}
    \caption{The experimental snapshots of U4-53C radial cut microstructures.}
    \label{fig:temp}
\end{figure}

\begin{figure}[!hbt]
    \centering
    \includegraphics[width=.6\linewidth]{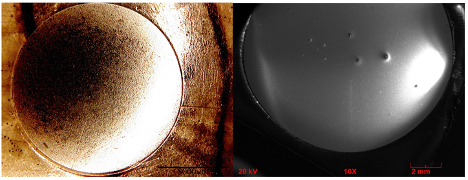}
    \caption{The experimental snapshots of U2-49 radial cut microstructures.}
    \label{fig:temp}
\end{figure}

\newpage
\subsubsection{Axial Fracture Images}\label{sec:axial}

\begin{figure}[!hbt]
    \centering
    \includegraphics[width=.6\linewidth]{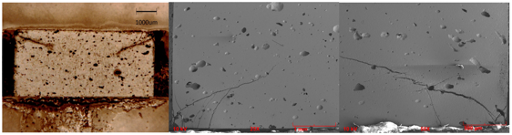}
    \caption{The experimental snapshots of U6-40 axial cut microstructures.}
    \label{fig:temp}
\end{figure}

\begin{figure}[!hbt]
    \centering
    \includegraphics[width=.6\linewidth]{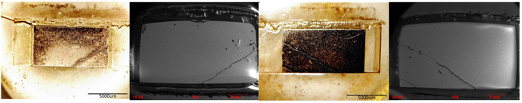}
    \caption{The experimental snapshots of U5-15A axial cut microstructures.}
    \label{fig:temp}
\end{figure}

\begin{figure}[!hbt]
    \centering
    \includegraphics[width=.6\linewidth]{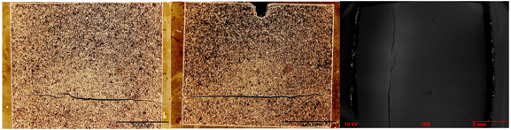}
    \caption{The experimental snapshots of U4-53C axial microstructures.}
    \label{fig:temp}
\end{figure}

\begin{figure}[!hbt]
    \centering
    \includegraphics[width=.6\linewidth]{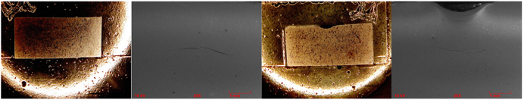}
    \caption{The experimental snapshots of U2-49 axial cut microstructures.}
    \label{fig:temp}
\end{figure}



\newpage
\bibliographystyle{elsarticle-num-names}
\bibliography{references.bib}







\end{document}